\documentclass[12pt,a4paper]{article}
\usepackage[utf8]{inputenc}
\usepackage{amsmath}
\usepackage{placeins}
\usepackage{url}
\usepackage{natbib}
\usepackage{color,xcolor,setspace,geometry}
\setcitestyle{authoryear,open={(},close={)}}
\usepackage{cite}
\usepackage{amsfonts}
\usepackage{pifont}

\usepackage{arydshln}
\usepackage{tikz}
\usetikzlibrary{shapes}
\usetikzlibrary{plotmarks}
\usetikzlibrary{tikzmark,matrix,calc}
\usetikzlibrary{arrows.meta,calc,decorations.markings,math,arrows.meta}
\usepackage{amssymb}
\usepackage{multirow}
\usepackage{graphicx}
\usepackage{soul}
\usepackage{rotating}
\usepackage{setspace}
\usepackage{threeparttable}
\usepackage{todonotes}
\usepackage{authblk}
\begin{document}
\begin{spacing}{1.35}	

\title{Bookmakers' mispricing of the disappeared home advantage in the German Bundesliga after the COVID-19 break}

\author[$\star$]{Christian Deutscher}
\author[$\star$ $\ddag$]{David Winkelmann}
\author[$\star$]{Marius Ötting}

\affil[$\star$]{Bielefeld University, Universit\"atsstrasse 25, Bielefeld, Germany.}
\affil[$\ddag$]{Corresponding author.}

\maketitle

\begin{abstract}
The outbreak of COVID-19 in March 2020 led to a shutdown of economic activities in Europe. This included the sports sector, since public gatherings were prohibited. The German Bundesliga was among the first sport leagues realising a restart without spectators. Several recent studies suggest that the home advantage of teams was eroded for the remaining matches. Our paper analyses the reaction by bookmakers to the disappearance of such home advantage. We show that bookmakers had problems to adjust the betting odds in accordance to the disappeared home advantage, opening opportunities for profitable betting strategies.
\end{abstract}

\newpage
\section{Introduction}
Playing in front of the home crowd is beneficial to team success in all of sports. Home fans use supportive chants to motivate their team and scream and shout at the opponent. Referees also appear to be influenced by the audience as their decisions have repeatedly shown to favour the home team \citep{dohmen2016referee}. These impacts can help to explain why chances to win a matchup are typically higher during home than away games. Such home advantage is also a well-known fact to bookmakers, who price in all information available and hence offer, on average, lower odds for bets on the home than on the away team. 

With the outbreak of the COVID-19 pandemic in spring 2020, all professional and amateur sports had to be cancelled because public gatherings were prohibited. About two months later, the German Bundesliga was among the first to resume playing, still in the same stadiums as before, but with fans absent. With games being played in empty stadiums, the home advantage eroded immediately \citep{fischer2020does, reade2020echoes}. While it took some rounds for this change to become pronounceable, it upheld until the season finished at the end of June 2020. With the German Bundesliga being a heavyweight in the betting market, the questions emerges whether bookmakers adapt their odds concerning the now non-existent home advantage.

This paper analyses betting market (in-)efficiencies in the German Bundesliga for matches behind closed doors after the COVID-19 break in spring 2020. Our analysis focuses on the 83 matches\footnote{In addition to rounds 26-34, two postponed matches were played behind closed doors.} at the end of the season 2019/2020 in the highest German football division and compares it to previous seasons. Our analyses show that bookmakers did not change their pricing during the period of matches after the start of no-attendance games. This opened highly profitable strategies to bettors with return on investment of around fifteen percent.

The paper is organised as follows: Section 2 provides an overview of the literature on the home advantage in sports as well as the home bias in sports betting. It is proceeded by descriptive statistics of our data in Section 3. Section 4 analyses the efficiency of betting markets before and after the COVID-19 break and strategies generating positive returns for bettors. Finally, Section 5 concludes with a discussion of our results.

\section{The home advantage and the home bias}
Our analysis considers the bookmakers' evaluation of the home advantage in football. We hence first present the state of research on the home advantage in football, to be followed by the incorporation of the home advantage into betting odds.

\subsection{The advantage of playing at home}
It is well documented in the literature that teams enjoy an advantage when playing at home. In football, teams have higher chances to win when playing at home than playing the exact team in an away game.\footnote{For meta-analysis non-exclusive to football see, e.g., \citet{courneya1992home, jamieson2010home}.} While the magnitude of such home advantage is more pronounced in football than in other sports \citep{jamieson2010home} and decreases to some extend over time \citep{palacios2004structural}, the exact mechanism of why teams have better winning records at home is still not well understood. 

Research agrees that it is not a singly source which generates the home advantage, but is rather assembled by a combination of different factors, which appear impossible to be disentangled empirically \citep{pollard2014components}. Since the away team has to travel, their journey as well as their unfamiliarity with the venue/stadium could reduce winning chances \citep{schwartz1977home, courneya1992home}. Concerning tactical formation, home teams tend towards a more offensive style of play compared to away teams, which could again benefit their chances of winning \citep{schwartz1977home, carmichael2005home}. Also, referees have shown to favour home teams, known as the \textit{home bias}. Such (unintentional) favouritism displays in longer extra-times when the home team is trailing compared to when the home team is winning \citep{sutter2004favoritism, garicano2005favoritism}.\footnote{See \citet{dohmen2016referee} for a review of the literature.}

The recent COVID-19 pandemic opened the opportunity to determine if the home advantage in football still exists when no spectators are present. The results are unanimous and find the home advantage to disappear once games were played behind closed doors. \citet{reade2020echoes} analyse various European football competitions including the German Bundesliga, while \citet{Dilger2020} and \citet{fischer2020does} focus on Germany only. The studies also agree that the referee bias towards the home teams dissapeared, potentially due to less social pressure of home fans on the referee (see also \citealp{endrich2020home}).

\subsection{(Miss-)Pricing of the home advantage by bookmakers}

Following the concept of efficient markets, asset prices (equivalent to betting odds) should contain all available information \citep{fama1970}. Since bookmakers face uncertainty of events and have to adjust to new information \citep{deutscher2018betting}, they keep a risk premium, referred to as margin. Such margins decreased in recent years due to increasing competition between bookmakers (\citealp{vstrumbelj2010online}). Efficient betting markets imply that market participants (bettors) cannot use simple strategies to beat the market and make profits, given the margin kept by the bookmaker. Those simple strategies include systematic betting on, e.g., home teams, favourites, popular or recently promoted teams.\footnote{\citet{winkelmann2020} present an overview of studies on biases in European betting markets.}

As the location of the game has shown to benefit the home team, the betting odds for home teams are on average lower than for away teams. The again so-called \textit{home bias} refers to increased payouts (equivalent to higher odds) for the home team compared to the fair odds. Such biased odds can arise from the bookmakers inability to predict game outcomes, their knowledge that fans are somewhat unable to assess teams' strength \citep{na2019not} or their goal to have a so-called balanced book, which is given if wagers on both outcomes (home and away win) level out such that the bookmakers secure a profit independent of the game outcome \citep{hodges2013fixed}. Since bettors rather bet on underdogs than favourites, bookmakers offer favourable odds on home wins \citep{franke2020market}. If such bias towards home wins is large enough to exceed the margin kept by the bookmaker, a profitable strategy would suggest to systematically bet on the home team. Supporting empirical evidence comes from e.g.\ \citet{forrest2008sentiment} and \citet{vlastakis2009efficient}. 

The COVID-19 pandemic offers an unique natural experiment: 
The direct effect on the disappearance of the home advantage in the German Bundesliga is strengthened by imperfect information of the bookmaker due to a small number of comparable games behind closed doors prior to the pandemic.\footnote{Prior to COVID-19} This leads to the question if and how fast bookmakers responded by adjusting the betting odds. From the bettors' perspective this opens the question if mispricing by bookmakers created profitable strategies.

\section{Data}
Due to spread of the COVID-19 pandemic in early 2020 the German Bundesliga prohibited the attendance of spectators after round 25 on March 9th 2020. By then 223 matches were played\footnote{25 rounds with nine games each; two games had been postponed due to other circumstances, one of them has been played on March 12th behind closed doors.} and 82 had been rescheduled between the 16th of May and the 27th of June 2020. The Bundesliga was the first of the top European football leagues to restart matches, but without spectators. Match data retrieved from \url{www.football-data.co.uk} cover results and pre-game betting odds for all games of the German Bundesliga from season 2014/15 until 2019/20. We focus on data covering the current season while using the five preceding seasons as reference to cover potential within season dynamics. We split the 2019/20 season into two periods, considering matches with and without fans. Previous seasons are separated after round 25 corresponding to the COVID-19 break in season 2019/20 for comparison.

\subsection{Descriptive match statistics}
Table \ref{tab:results} provides an overview on match results. Prior to the most recent season, home teams won nearly half of their matches, with an almost equal split between away wins (29.42\%) and draws (25.60\%). At the end of a season the number of home wins increases by more than three percentage points while games resulting in a draw and away wins are less likely than before.
The 2019/20 German Bundesliga season stands out due to the large number of away wins which is nearly 20\% higher compared to previous seasons even in rounds with spectators. The proportion of home wins and draws is smaller by 2 and about 3.5 percentage points, respectively. While the proportion of draws increases slightly after the COVID-19 break and equals the amount at the end of the season in previous years, we find a strong increase in the number of away wins, which are more than 25\% higher than in previous rounds and 65\% higher than at the end of previous seasons. At the same time the number of home wins decreases and is about 25\% lower than before the COVID-19 break and about 35\% lower than in previous seasons (see Table \ref{tab:results}). These findings confirm previous results which revealed an eroded home effect for the German Bundesliga after the COVID-19 break (see e.g.\ \citealp{Dilger2020}).

\begin{table}[h]
    \centering
    \scalebox{0.97}{
    \begin{tabular}{c|c|ccc}
         & Matches & Home wins & Draws & Away wins \\
         \hline
        Seasons 2014/15-2018/19 Round 1-25 & 1125 & 44.98\% & 25.60\% & 29.42\% \\
        Seasons 2014/15-2018/19 Round 26-34 & 405 & 49.63\% & 23.46\% & 26.91\%\\
        Season 2019/20 with spectators & 223 & 43.05\% & 21.97\% & 34.98\% \\
        Season 2019/20 without spectators & 83 & 32.53\% & 22.89\% & 44.58\%\\
    \end{tabular}}
    \caption{Proportion of match outcomes before and after the COVID-19 break as well as for previous seasons.}
    \label{tab:results}
\end{table}

Similar results occur regarding the average number of goals scored by the home and away team. In previous years, the number of total goals increased at the end of the season from 2.82 goals on average to 3.06 goals. However, in 2019/20 considerably more goals were already scored in games with spectators, which mainly origins from the increased number of away goals. While away teams scored even more goals after the COVID-19 break, the number of home goals decreased by nearly 20\%. For the first time, the number of away goals surpasses the number of home goals (see Table \ref{tab:goals}).

\begin{table}[h]
    \centering
    \scalebox{0.97}{
    \begin{tabular}{c|cc|c}
         & Home goals & Away goals & Total goals \\
         \hline
        Seasons 2014/15-2018/19 Round 1-25 & 1.58 & 1.24 & 2.82 \\
        Seasons 2014/15-2018/19 Round 26-34 & 1.80 & 1.26 & 3.06 \\
        Season 2019/20 with spectators & 1.74 & 1.51 & 3.25 \\
        Season 2019/20 without spectators & 1.43 & 1.66 & 3.10 \\
    \end{tabular}}
    \caption{Average home and away goals before and after the COVID-19 break as well as for previous seasons.}
    \label{tab:goals}
\end{table}

\subsection{Implied winning probabilities and bookmaker's margins}
To determine the precision of bookmakers' forecasts, we first turn our attention to the betting odds. As betting odds contain a margin, they have to be adjusted as follows to obtain the implied winning probability $\hat{\pi}_i$ given by the bookmaker (see e.g.\ \citealp{deutscher2013sabotage}):
$$
\hat{\pi}_{i}=\frac{1/O_i}{1/O_h+1/O_d+1/O_a}, \,\,\,\, i = h,d,a
$$
where $O_i$ represents the average odds over all bookmakers for home wins ($i=h$), away wins ($i=a$), and draws ($i=d$).\footnote{The data set covers odds of between 30 and 56 bookmakers for each match. The pairwise correlation between betting odds offered by different bookmakers is at least 0.969 for home wins and 0.945 for away wins.} The difference in the implied probabilities $\textit{ImpProbDiff}=\hat{\pi}_h-\hat{\pi}_a$ indicates whether the bookmaker denotes the home team to be the favourite ($\textit{ImpProbDiff}>0$), whereas $\textit{ImpProbDiff}<0$ coincides with a favoured away team. If the (absolute) difference in the implied winning probability between two teams in a specific match is large, one team can clearly declared to be the favourite. In contrast, a small difference indicates that the bookmaker assigns nearly equal abilities to both teams.

Bookmakers use their margin to account for possible mispricing and to remain profitable. Margins for each match are calculated as  $\sum\limits_{i\in\{h,d,a\}}O_{m,i}^{-1}-1$ for matches $m=1,\ldots,M$. Table \ref{tab:margins} compares margins for season 2019/20 to previous seasons, indicating that margins are lower in the most recent season. This confirms previous results and is in line with decreased margins over time caused by higher market competition as argued by \citet{forrest2005odds} and \citet{vstrumbelj2010online}. 

\begin{table}[h]
    \centering
    \begin{tabular}{c|cc|c}
         & Margins \\
         \hline
        Seasons 2014/15-2018/19 Round 1-25 & 5.09\% \\
        Seasons 2014/15-2018/19 Round 26-34 & 5.11\% \\
        Season 2019/20 with spectators & 4.83\% \\
        Season 2019/20 without spectators & 4.79\% \\ 
    \end{tabular}
    \caption{Bookmaker's margins before and after the COVID-19 break as well as for previous seasons.}
    \label{tab:margins}
\end{table}

Furthermore, margins typically increase as the assessment of teams becomes more difficult, e.g.\ if the bookmaker expect two teams to have nearly equal abilities. Table \ref{tab:regmar} displays results of a regression model where we explain the margin for a specific match by the absolute difference between the implied winning probability for a home and away win as well as the season. We find significantly lower margins if this difference increases, i.e.\ if one team can clearly declared to be the favourite.

\begin{table}[!htbp] \centering 
  \caption{Margins explained by absolute difference in implied winning probabilities, season and round.} 
  \label{tab:regmar} 
  \scalebox{0.8}{
\begin{tabular}{@{\extracolsep{5pt}}lc} 
\\[-1.8ex]\hline 
\hline \\[-1.8ex] 
 & \multicolumn{1}{c}{\textit{Dependent variable:}} \\ 
\cline{2-2} 
\\[-1.8ex] & Margin \\ 
\hline \\[-1.8ex] 
 \textit{Absolute difference in implied probabilities} & $-$0.002$^{***}$ \\ 
  & (0.0003) \\ 
  & \\ 
 \textit{Season} & $-$0.001$^{***}$ \\ 
  & (0.00004) \\ 
  & \\ 
 \textit{Constant} & 0.056$^{***}$ \\ 
  & (0.0002) \\ 
  & \\ 
\hline \\[-1.8ex] 
Observations & 1,836 \\ 
R$^{2}$ & 0.468 \\ 
\hline 
\hline \\[-1.8ex] 
\textit{Note:}  & \multicolumn{1}{r}{$^{*}$p$<$0.1; $^{**}$p$<$0.05; $^{***}$p$<$0.01} \\ 
\end{tabular}} 
\end{table} 

However, according to Table \ref{tab:margins}, there is only a slight increase in the margins at the end of previous seasons while margins slightly decrease after the COVID-19 break. While increased margins at the end of previous seasons can be referred to the difficulty in the prediction of match outcomes when there are matches with two teams without any possibilities to face promotion to the international competition or are under threat of possible relegation, findings on season 2019/20 are somewhat surprising. For matches behind closed doors, previous experience of bookmakers is fairly low. Therefore the assessment of teams is expected to be difficult. It could be expected that bookmakers account for this by increasing margins.

\section{Analysing market inefficiencies}
Bookmakers had close to no experience on games behind closed doors, since only a small number of such games had been played in the past. Nevertheless, as indicated by our findings from the previous section, they did not account for this increased uncertainty by increasing their margins. Our descriptive analysis also reveals a disappeared home advantage, equal to more away wins after the COVID-19 break at the end of season 2019/20 in the German Bundesliga. This opens the question whether bookmakers incorporated increased winning probabilities for away teams in their odds. Otherwise, bettors would have been able to gain positive returns when betting consistently on the away team. In accordance to the home bias as described above, this phenomenon can be denoted as an away bias. In the following, we provide regression analyses to statistically test for increased chances to win a bet when betting on away teams (especially in games played behind closed doors).

\subsection{Model}
Our analysis covers bets on home and away teams since margins for draws only vary slightly in football \citep{pope1989information}. Due to the differences in the winning probability when playing at home or away we include the binary variable \textit{Away} indicating whether we bet on the away team (\textit{Away=1}) or the home team (\textit{Away=0}). In addition, the variable \textit{COVID-19} equals one if the match has been played behind closed doors. To control for possible dynamic changes during the course of the season the variable $\textit{Betting after round 25}$ equals one if we bet after this round. We further include an interaction term between \textit{Away} and \textit{COVID-19}. This term allows for differences in the away bias between matches with and without spectators. In the second model, we additionally test for adjustments by the bookmaker during the period of matches without spectators, thus including the \textit{Round after round 25} taking value 1 for round 26 up to value 9 for round 34 as well as an interaction term with the \textit{COVID-19} variable.

Efficient markets would imply that neither betting on away teams nor the \textit{COVID-19} variable affects the chance to win a bet significantly. To test if the market considered here is efficient, we run a logistic regression model to detect whether any variable beyond the implied probability has explanatory power on the dependent binary variable which indicates if a bet was \textit{Won}. The linear predictor with $\text{logit}(\Pr(Won_i = 1)) = \eta_i$ is linked by the logit function and given as:
\begin{equation*}
\begin{split}
    \eta_i &=\beta_0+\beta_1 \text{\textit{Implied Probability}}_i+\beta_2 \text{\textit{Away}}_i + \beta_3 \text{\textit{Betting after round 25}}_i\\
    &+ \beta_4 \text{\textit{COVID-19}}_i + \beta_5 \text{\textit{Away}}_i \cdot \text{\textit{COVID-19}}_i + \beta_6 \text{\textit{Round after round 25}}_i\\
    &+ \beta_7 \text{\textit{Round after round 25}}_i \cdot \text{\textit{COVID-19}}_i.
\end{split}
\end{equation*}
We use maximum likelihood to fit the models to the data using the function \texttt{glm()} in R \citep{R}. This approach follows the concept of various further studies as \citet{forrest2008sentiment}, \citet{franck2011sentimental}, and \citet{feddersen2017sentiment}.

\subsection{Results}
Table \ref{tab:regmod} displays the estimated coefficients and standard errors of our two regression models. The \textit{Implied Probability} calculated based on bookmakers' odds has strong explanatory power on the actual outcome of the game, which is intuitively plausible. The negative and significant effect of \textit{Away} reveals a home bias for matches with spectators in the German Bundesliga which is in line with the existing literature. However, there is no significant change in the chances to win a bet in the last quarter of the season as indicated by the insignificant effect of the dummy variable for \textit{Betting after round 25}. Model 1 allows for a comparison between a bet on a match without spectators and a match at the end of a previous season with equal implied probability by the bookmaker. Here we find a considerable decrease in the odds to win a bet on the home team while the odds to win a bet on the away team are considerably increased.

Model 2 additionally tests for an adjustment of betting odds by the bookmaker during the period of matches behind closed doors. We do not find significant changes in the chances to win a bet for subsequent rounds, thus concluding that the misassessment is not limited to the first rounds after the restart of the German Bundesliga but kept up for the remainder of the season.

\begin{table}[!htbp] \centering 
  \caption{Estimation results of our model fitted to the whole data set.} 
  \label{tab:regmod} 
  \scalebox{0.4}{
\begin{tabular}{@{\extracolsep{5pt}}lcc} 
\\[-1.8ex]\hline 
\hline \\[-1.8ex] 
 & \multicolumn{2}{c}{\textit{Dependent variable:}} \\ 
\cline{2-3} 
\\[-1.8ex] & \multicolumn{2}{c}{Won} \\ 
\\[-1.8ex] & Model 1 & Model 2\\ 
\hline \\[-1.8ex] 
 \textit{Implied probability} & 4.530$^{***}$ & 4.530$^{***}$ \\ 
  & (0.231) & (0.231) \\ 
  & & \\ 
 \textit{Away} & $-$0.162$^{**}$ & $-$0.162$^{**}$ \\ 
  & (0.080) & (0.080) \\ 
  & & \\ 
 \textit{Betting after round 25} & 0.032 & 0.076 \\ 
  & (0.089) & (0.174) \\ 
  & & \\ 
 \textit{COVID-19} & $-$0.606$^{**}$ & $-$0.916$^{**}$ \\ 
  & (0.268) & (0.445) \\ 
  & & \\ 
 \textit{Away} $\cdot$ \textit{COVID-19} & 1.136$^{***}$ & 1.143$^{***}$ \\ 
  & (0.358) & (0.359) \\ 
  & & \\ 
 \textit{Round after round 25} &  & $-$0.009 \\ 
  &  & (0.030) \\ 
  & & \\ 
 \textit{Round after round 25} $\cdot$ \textit{COVID-19} &  & 0.063 \\ 
  &  & (0.071) \\ 
  & & \\ 
 \textit{Constant} & $-$2.207$^{***}$ & $-$2.207$^{***}$ \\ 
  & (0.118) & (0.118) \\ 
  & & \\ 
\hline \\[-1.8ex] 
Observations & 3,672 & 3,672 \\ 
Akaike Inf. Crit. & 4,322.426 & 4,325.643 \\ 
\hline 
\hline \\[-1.8ex] 
\textit{Note:}  & \multicolumn{2}{r}{$^{*}$p$<$0.1; $^{**}$p$<$0.05; $^{***}$p$<$0.01} \\ \end{tabular}}
\end{table}

\begin{figure}[!htb]
\centering
\includegraphics[scale=.7]{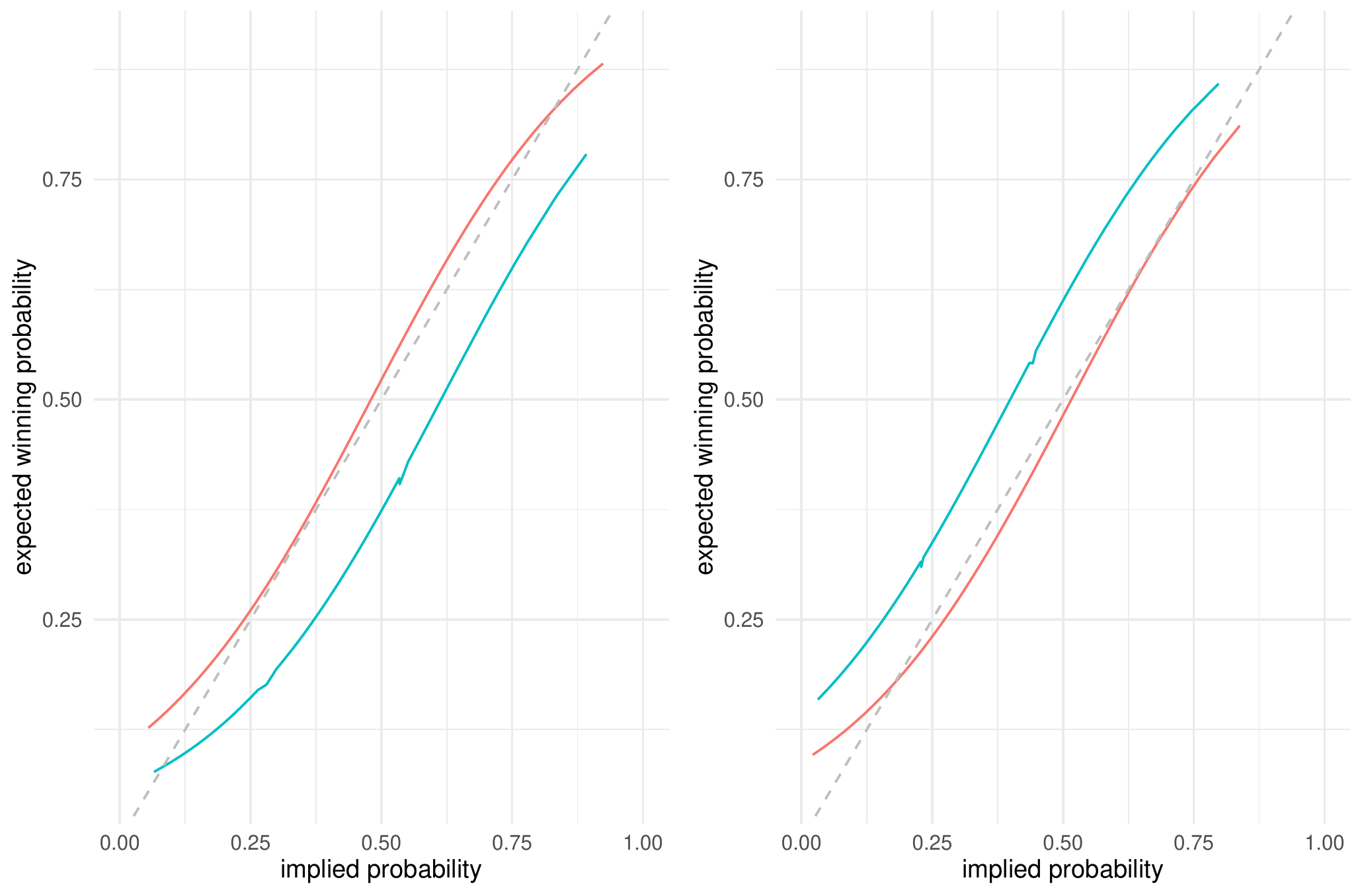}
\caption{Comparison between implied probabilities by the bookmaker and expected winning probability under the model for home teams (left panel) and away teams (right panel) for matches at the end of previous seasons (red) and matches behind closed doors (blue).} 
\label{fig:ImpProb}
\end{figure}

We use the results in model 1 to compare implied probabilities given by the bookmaker to winning probabilities as expected under our model (see Figure \ref{fig:ImpProb}). The left panel represents bets on home teams while bets on the away team are illustrated in the right panel. Bets at the end of a previous season (with spectators) are depicted in red while bets on a game behind closed doors are represented in blue colour. An efficient market would correspond to the dashed diagonal line. We find expected winning probabilities to deviate only slightly from the efficient market line for matches with spectators at the end of previous seasons. As indicated by our model, the figure confirms significantly higher expected probabilities to win a bet when betting on away teams after the COVID-19 break, while these chance are significantly lower for bets on home teams (see Figure \ref{fig:ImpProb}). This underlines that bookmakers were unable to incorporate the disappearing home advantage for matches without the attendance of spectators into their odds, potentially opening the possibility for positive returns to bettors who systematically bet on away wins.

\subsection{Returns}
Since the bookmakers take a margin, significant impact of variables aside from the implied probabilities does not necessarily turn into profitable strategies. Still, our model indicates significantly higher chances to win a bet when betting on away games after the COVID-19 break. We use this result to calculate return on investments (ROIs) for simple strategies, i.e.\ betting consistently on home or away teams. For previous seasons we find on average higher returns when betting on the home team (see Table \ref{tab:returns}). This confirms the significant negative effect of the \textit{Away} variable in our regression model. However, until round 25 the average return is still negative due to the bookmakers' margin. The increased number of home wins at the end of previous seasons (see Table \ref{tab:results}) leads to positive returns of about 6.24\% when betting on the home team at this time (Table \ref{tab:returns}).

Even if our regression model indicates a home bias over all matches between 2014 and round 25 of season 2019/20, the descriptive analysis shows a considerably higher proportion of away wins also at the beginning of season 2019/20 (see Table \ref{tab:results}). Bookmakers were not able to include this shift into their odds leading to positive returns of about 5.53\% when betting on away teams. As the home advantage disappears after the COVID-19 pandamic leading to even more away wins, positive and considerable returns of nearly 15\% can be generated with this strategy in matches behind closed doors. Meanwhile, betting on home teams generates an average return of -33.84\% (Table \ref{tab:returns}).

\begin{table}[h]
    \centering
    \begin{tabular}{c|cc}
         & Betting on home win & Betting on away win \\
         \hline
        Seasons 2014/15-2018/19 Round 1-25 & -1.37\% & -11.69\% \\
        Seasons 2014/15-2018/19 Round 26-34 & 6.24\% & -15.52\% \\
        Season 2019/20 with spectators & -6.64\% & 5.53\% \\
        Season 2019/20 without spectators & -33.84\% & 14.71\% \\ 
    \end{tabular}
    \caption{Returns on investment when betting on the home and away team before and after the COVID-19 break as well as for previous seasons.}
    \label{tab:returns}
\end{table}

The results open the question whether these considerable returns are mainly driven by a few bets with high odds, i.e.\ wins of away teams which can be clearly denoted to be the underdog, or several wins of teams with low odds. As stated above, we denote the difference in percentage points between the implied probabilities given by the bookmaker for a home and away win by \textit{ImpProbDiff}. While positive values of this variable indicate higher implied probabilities for the home team, \textit{ImpProbDiff} takes negative values if the bookmaker denote the away team to be the favourite. Table \ref{tab:odds} covers different ranges for this variable together with the number of matches, home wins, draws, and away wins in this group.

\begin{table}[h] \centering 
    \begin{tabular}{c|cccc}
        ImpProbDiff & Matches & Home wins & Draws & Away wins \\ 
        \hline \\[-1.8ex] 
        Heavy home favourite & & & & \\
        $[\hspace{0.345cm} 0.90;\hspace{0.345cm} 0.75)$ & $3$ & $2$ & $1$ & $0$ \\ 
        $[\hspace{0.345cm} 0.75;\hspace{0.345cm} 0.60)$ & $6$ & $3$ & $2$ & $1$ \\ 
        $[\hspace{0.345cm} 0.60;\hspace{0.345cm} 0.45)$ & $7$ & $6$ & $1$ & $0$ \\ 
        $[\hspace{0.345cm} 0.45;\hspace{0.345cm} 0.30)$ & $9$ & $5$ & $2$ & $2$ \\ 
        $[\hspace{0.345cm} 0.30;\hspace{0.345cm} 0.15)$ & $12$ & $3$ & $4$ & $5$ \\ 
        $[\hspace{0.345cm} 0.15;\hspace{0.345cm} 0.00)$ & $13$ & $4$ & $3$ & $6$ \\ 
        Balanced match & & & & \\
        $[\hspace{0.345cm} 0.00;-0.15)$ & $10$ & $1$ & $4$ & $5$ \\ 
        $[-0.15;-0.30)$ & $4$ & $1$ & $0$ & $3$ \\ 
        $[-0.30;-0.45)$ & $9$ & $2$ & $1$ & $6$ \\ 
        $[-0.45;-0.60)$ & $6$ & $0$ & $1$ & $5$ \\ 
        $[-0.60;-0.75]$ & $4$ & $0$ & $0$ & $4$ \\ 
        Heavy away favourite & & & & \\
        \hline \\[-1.8ex] 
    \end{tabular} 
    \caption{Match outcomes depending on difference in implied winning probabilities given by the bookmaker.} 
    \label{tab:odds} 
\end{table} 

In about 60\% of the matches behind closed doors in the German Bundesliga, the home team was denoted to be the favourite according to the bookmaker, i.e.\ implied winning probabilities for the home team were higher than for the away team (see Table \ref{tab:odds}).\footnote{On average, the home team had implied winning probabilities exceeding the value of the away team by 7.73 percentage points.} Considering matches with a large difference of more than 45 percentage points, we find that only 11 of 16 such heavy home favourites could win their game while nine out of ten away teams with considerably larger implied winning probabilities won their game. Focusing on 39 relatively close matches, i.e.\ those where the difference in the implied probability did not exceed 0.3, we find 19 away in contrast to only 9 home wins. Even if the bookmaker denoted the home team to be a slight favourite, we find 11 away wins compared to only seven home wins in these matches (see Table \ref{tab:odds}). Accordingly, bookmakers undervalue the strength of away teams for matches behind closed doors, especially for balanced competitions.

\section{Discussion}
We analyse the home advantage and its evaluation by the bookmakers in Bundesliga games that were played behind closed doors. While the number of home wins increased at the end of previous seasons, it disappeared during the COVID-19 seasons, equal to considerably more away wins at the end of season 2019/20. Furthermore, the number of home goals decreased as the number of away goals increased,  confirming the disappearance of the home advantage. Our analysis shows the bookmakers' struggle to price such change. Their odds imply that the home advantage remained intact and opened opportunities for betters to generate substantial profits of around 15\% when betting on away teams in games that were played behind closed doors. A more fine-grained analysis shows that especially in close competitions away teams won considerably more games than expected by the bookmakers, confirming their struggle to implement the impact of games without spectators on the home advantage. Returns of presented magnitude are very rare in betting markets and can have significant impact on bookmakers business as well as other stakeholders. Since many Bundesliga teams are sponsored by bookmakers, they also rely on efficient betting odds and profitable business for the bookmaker. Given the high turnover in football betting, the overall impact of inefficient markets is considerably. 

While we clearly indicate that bookmakers were not able to incorporate the disappearance of the home advantage for matches behind closed doors in the German Bundesliga into their betting odds it remains unclear why there the home advantage somewhat disappeared in the 2019/20 season prior to COVID-19. Furthermore, the magnitude of the home advantage and, therefore, also the decrease may not be the same for all teams, but depend on the team's popularity or the stadium capacity. Thus, it would be interesting to extend the analysis to further top European football leagues as the Englisch Premier League, the Spanish La Liga or the Italian Serie A after their restart with matches behind closed doors some weeks after the German Bundesliga. Finally, it is expected that spectators will be able to return into the stadium in autumn with strong restrictions. This opens the possibility for a comparison between matches behind closed doors, a reduced stadium capacity and the full support of spectators.
\newpage

\bibliographystyle{apalike}
\bibliography{refs}

\begin{thebibliography}{}

\bibitem[Carmichael and Thomas, 2005]{carmichael2005home}
Carmichael, F. and Thomas, D. (2005).
\newblock Home-field effect and team performance: evidence from english
  premiership football.
\newblock {\em Journal of sports economics}, 6(3):264--281.

\bibitem[Courneya and Carron, 1992]{courneya1992home}
Courneya, K.~S. and Carron, A.~V. (1992).
\newblock The home advantage in sport competitions: a literature review.
\newblock {\em Journal of Sport \& Exercise Psychology}, 14(1).

\bibitem[Deutscher et~al., 2013]{deutscher2013sabotage}
Deutscher, C., Frick, B., G{\"u}rtler, O., and Prinz, J. (2013).
\newblock Sabotage in tournaments with heterogeneous contestants: Empirical
  evidence from the soccer pitch.
\newblock {\em The Scandinavian Journal of Economics}, 115(4):1138--1157.

\bibitem[Deutscher et~al., 2018]{deutscher2018betting}
Deutscher, C., Frick, B., and {\"O}tting, M. (2018).
\newblock Betting market inefficiencies are short-lived in {G}erman
  professional football.
\newblock {\em Applied Economics}, 50(30):3240--3246.

\bibitem[Dilger and Vischer, 2020]{Dilger2020}
Dilger, A. and Vischer, L. (2020).
\newblock No home bias in ghost games.
\newblock {\em Discussion Paper of the Institute for Organisational Economics}.

\bibitem[Dohmen and Sauermann, 2016]{dohmen2016referee}
Dohmen, T. and Sauermann, J. (2016).
\newblock Referee bias.
\newblock {\em Journal of Economic Surveys}, 30(4):679--695.

\bibitem[Endrich and Gesche, 2020]{endrich2020home}
Endrich, M. and Gesche, T. (2020).
\newblock Home-bias in referee decisions: Evidence from'ghost matches' during
  the covid-19 pandemic.
\newblock {\em Center for Law \& Economics Working Paper Series}.

\bibitem[Fama, 1970]{fama1970}
Fama, E.~F. (1970).
\newblock Efficient capital markets: a review of theory and empirical work.
\newblock {\em The Journal of Finance}, 25(2):383--417.

\bibitem[Feddersen et~al., 2017]{feddersen2017sentiment}
Feddersen, A., Humphreys, B.~R., and Soebbing, B.~P. (2017).
\newblock Sentiment bias and asset prices: evidence from sports betting markets
  and social media.
\newblock {\em Economic Inquiry}, 55(2):1119--1129.

\bibitem[Fischer and Haucap, 2020]{fischer2020does}
Fischer, K. and Haucap, J. (2020).
\newblock Does crowd support drive the home advantage in professional soccer?
  evidence from german ghost games during the covid-19 pandemic.
\newblock Technical report, DICE Discussion Paper.

\bibitem[Forrest et~al., 2005]{forrest2005odds}
Forrest, D., Goddard, J., and Simmons, R. (2005).
\newblock Odds-setters as forecasters: the case of {E}nglish football.
\newblock {\em International Journal of Forecasting}, 21(3):51--564.

\bibitem[Forrest and Simmons, 2008]{forrest2008sentiment}
Forrest, D. and Simmons, R. (2008).
\newblock Sentiment in the betting market on {S}panish football.
\newblock {\em Applied Economics}, 40(1):119--126.

\bibitem[Franck et~al., 2011]{franck2011sentimental}
Franck, E., Verbeek, E., and N{\"u}esch, S. (2011).
\newblock Sentimental preferences and the organizational regime of betting
  markets.
\newblock {\em Southern Economic Journal}, 78(2):502--518.

\bibitem[Franke, 2020]{franke2020market}
Franke, M. (2020).
\newblock Do market participants misprice lottery-type assets? {E}vidence from
  the {E}uropean soccer betting market.
\newblock {\em The Quarterly Review of Economics and Finance}, 75(1):1--18.

\bibitem[Garicano et~al., 2005]{garicano2005favoritism}
Garicano, L., Palacios-Huerta, I., and Prendergast, C. (2005).
\newblock Favoritism under social pressure.
\newblock {\em Review of Economics and Statistics}, 87(2):208--216.

\bibitem[Hodges et~al., 2013]{hodges2013fixed}
Hodges, S., Lin, H., and Liu, L. (2013).
\newblock Fixed odds bookmaking with stochastic betting demands.
\newblock {\em European Financial Management}, 19(2):399--417.

\bibitem[Jamieson, 2010]{jamieson2010home}
Jamieson, J.~P. (2010).
\newblock The home field advantage in athletics: A meta-analysis.
\newblock {\em Journal of Applied Social Psychology}, 40(7):1819--1848.

\bibitem[Na et~al., 2019]{na2019not}
Na, S., Su, Y., and Kunkel, T. (2019).
\newblock Do not bet on your favourite football team: the influence of fan
  identity-based biases and sport context knowledge on game prediction
  accuracy.
\newblock {\em European Sport Management Quarterly}, 19(3):396--418.

\bibitem[Palacios-Huerta, 2004]{palacios2004structural}
Palacios-Huerta, I. (2004).
\newblock Structural changes during a century of the world’s most popular
  sport.
\newblock {\em Statistical Methods and Applications}, 13(2):241--258.

\bibitem[Pollard and G{\'o}mez, 2014]{pollard2014components}
Pollard, R. and G{\'o}mez, M.~A. (2014).
\newblock Components of home advantage in 157 national soccer leagues
  worldwide.
\newblock {\em International Journal of Sport and Exercise Psychology},
  12(3):218--233.

\bibitem[Pope and Peel, 1989]{pope1989information}
Pope, P.~F. and Peel, D.~A. (1989).
\newblock Information, prices and efficiency in a fixed-odds betting market.
\newblock {\em Economica}, 56(223):323--341.

\bibitem[{R Core Team}, 2019]{R}
{R Core Team} (2019).
\newblock {\em R: A Language and Environment for Statistical Computing}.
\newblock R Foundation for Statistical Computing, Vienna, Austria.

\bibitem[Reade et~al., 2020]{reade2020echoes}
Reade, J.~J., Schreyer, D., and Singleton, C. (2020).
\newblock Echoes: what happens when football is played behind closed doors?
\newblock {\em Available at SSRN 3630130}.

\bibitem[Schwartz and Barsky, 1977]{schwartz1977home}
Schwartz, B. and Barsky, S.~F. (1977).
\newblock The home advantage.
\newblock {\em Social forces}, 55(3):641--661.

\bibitem[{\v{S}}trumbelj and {\v{S}}ikonja, 2010]{vstrumbelj2010online}
{\v{S}}trumbelj, E. and {\v{S}}ikonja, M.~R. (2010).
\newblock Online bookmakers’ odds as forecasts: the case of {E}uropean soccer
  leagues.
\newblock {\em International Journal of Forecasting}, 26(3):482--488.

\bibitem[Sutter and Kocher, 2004]{sutter2004favoritism}
Sutter, M. and Kocher, M.~G. (2004).
\newblock Favoritism of agents--the case of referees' home bias.
\newblock {\em Journal of Economic Psychology}, 25(4):461--469.

\bibitem[Vlastakis et~al., 2009]{vlastakis2009efficient}
Vlastakis, N., Dotsis, G., and Markellos, R.~N. (2009).
\newblock How efficient is the european football betting market? {E}vidence
  from arbitrage and trading strategies.
\newblock {\em Journal of Forecasting}, 28(5):426--444.

\bibitem[Winkelmann et~al., 2020]{winkelmann2020}
Winkelmann, D., Ötting, M., and Deutscher, C. (2020).
\newblock Short and long-term biases in european football pre-game betting
  markets.
\newblock {\em Universität Bielefeld Working Papers in Economics and
  Management}, 2020(06).

\end{thebibliography}

\end{spacing}
\end{document}